\documentclass[11pt]{article}
\begin{document}

\begin{center}
{\bf \huge Characterising exoplanets and their environment with UV transmission spectroscopy}
\end{center}

\smallskip

\noindent L. Fossati$^{1}$, V. Bourrier$^{2}$, D. Ehrenreich$^{2}$, C.\,A. Haswell$^{3}$, K.\,G. Kislyakova$^{4}$, H. Lammer$^{4}$, A. Lecavelier des Etangs$^{5}$, Y. Alibert$^{6}$, T.\,R. Ayres$^{7}$, G.\,E. Ballester$^{8}$, J. Barnes$^{3}$, D.\,V. Bisikalo$^{9}$, A. Collier, Cameron$^{10}$, S. Czesla$^{11}$, J.-M. D{\'e}sert$^{7}$, K. France$^{7}$, M. G\"udel$^{12}$, E. Guenther$^{13}$, Ch. Helling$^{10}$, K. Heng$^{6}$, M. Homstr\"om$^{14}$, L. Kaltenegger$^{15}$, T. Koskinen$^{8}$, A.\,F. Lanza$^{16}$, J.\,L. Linsky$^{17}$, C. Mordasini$^{6}$, I. Pagano$^{16}$, D. Pollacco$^{18}$, H. Rauer$^{19}$, A. Reiners$^{20}$, M. Salz$^{11}$, P.\,C. Schneider$^{21}$, V.\,I. Shematovich$^{9}$, D. Staab$^{3}$, A.\,A. Vidotto$^{2}$, P.\,J. Wheatley$^{18}$, B.\,E. Wood$^{22}$, R.\,V. Yelle$^{8}$\\

\vspace{-0.5cm}
\begin{scriptsize}
\noindent 
1.\,Argelander Institut f\"ur Astronomie, Universit\"at Bonn, Germany, lfossati$@$astro.uni-bonn.de; 
2.\,Observatoire astronomique de l'Universit\'e de Gen\`eve, Switzerland;  
3.\,Open University, UK;  
4.\,Space Research Institute, Graz, Austria;  
5.\,Institut d'astrophysique de Paris, France; 
6.\,University of Bern, Switzerland;
7.\,CASA, University of Colorado, USA;
8.\,LPL, University of Arizona, USA;
9.\,INASAN, Moscow, Russia;
10.\,University of St. Andrews, UK;
11.\,Universit\"at Hamburg, Germany;
12.\,University of Vienna, Austria;
13.\,TLS Tautenburg, Germany;
14.\,Swedish Institute of Space Physics, Sweden;
15.\,MPIA, Heidelberg, Germany;
16.\,INAF-Catania, Italy;
17.\,JILA, University of Colorado, USA;
18.\,University of Warwick, UK;
19.\,TU Berlin, Germany;
20.\,Universit\"at G\"ottingen, Germany;
21.\,ESA, The Netherlands;
22.\,US NRL, Washington, D.C., USA;
\end{scriptsize}

\bigskip

{\bf \large \noindent Introduction}

\smallskip

\noindent Extra-solar planets (exoplanets) have provided many fascinating surprises and puzzles since their discovery. Given the youth of exoplanet science, the field is strongly driven by observations: the Hubble Space Telescope (HST) is providing fundamental observations for the advance of exoplanet science, particularly in terms of atmospheric characterisation, crucial for our further understanding of planet formation and evolution.

The very first detection of an exoplanet atmosphere was via STIS transmission spectroscopy of the close-in giant planet (hot-Jupiter) HD\,209458\,b that led to the detection of sodium in the planet atmosphere [1]. Following this seminal detection, Vidal-Madjar et al.\,(2003;\,[2]) used STIS far-ultraviolet (far-UV) Ly\,$\alpha$ transmission spectroscopy of the same system to reveal that the planet possesses an extended hydrogen atmosphere that is escaping hydrodynamically [3]. This result led to the full realisation that atmospheric evaporation is a key factor shaping planet evolution (e.g., [4]).

Evaporation has also an impact on our understanding of planet formation and population. Atmospheric escape must be taken into account in order to use Gyr old planets to understand planet formation (e.g., the efficiency of accretion of primordial H/He atmospheres). High mass-loss rates would imply the existence of a consistent population of almost atmosphereless low mass close-in planets (e.g., CoRoT-7\,b and Kepler-10\,b), where some of these might be the remnant core of evaporated Neptune-mass planets [5,6,7]. High mass-loss rates would also imply that close-in planets may be engulfed by the host star [8] or become undetectable, leaving behind a gap in the observed planet population distribution [9,10,11].

The mechanism of evaporation, the upper atmosphere (exosphere) of exoplanets, and its interactions with the host star can be thoroughly studied \emph{only} at UV wavelengths and HST is the \emph{only} facility capable to collect the necessary observations. HST has now a limited lifetime and the next major UV space telescope, and still not as powerful as HST, might be able to fly at best only in 10 years time, or more (e.g., WSO-UV\footnote{\tt http://www.wso-uv.org/wso-uv2/index.php?lang=en}, ARAGO\footnote{\tt http://lesia.obspm.fr/UVMag/}). 

It is only in the last few years that a consistent number of planets orbiting stars bright enough for UV transit observations with HST ($V$\,$<$\,10\,mag) has been discovered by ground-based surveys (e.g., SuperWASP, HATNet, KELT), and many more are expected to be found in the next couple of years from the ground and by the CHEOPS and TESS space missions. It is therefore now time to use the exceptional capabilities of HST to create a treasury data archive of UV transit observations for planet atmospheric characterisation, allowing the whole community to bring forward this fundamental branch of exoplanet research for the next decade, and more.

\bigskip

{\bf \large \noindent Current status of research}

\smallskip

\noindent To date, only three hot-Jupiters have been thoroughly observed at UV wavelengths. HD\,209458\,b, with multiple transits observed at both far-UV and near-UV wavelengths, is a bloated (i.e., low density) hot-Jupiter orbiting a Sun-like star. HD\,189733\,b, with multiple transits observed only in the far-UV, is a hot-Jupiter orbiting a very active K-type star. WASP-12\,b, with multiple transits observed only in the near-UV, is a bloated hot-Jupiter orbiting a late F-type star and one of the hottest planets known to date\footnote{Single-transit far-UV observations exist also for the 55\,Cnc [12] and Gliese\,436 [13] systems, but the results require further confirmation.}.  

The common result obtained from the analysis of the UV data collected for these three planets is the presence of an extended, evaporating upper atmosphere [2,14,15]. This so-called hydrodynamic ``blow-off'' sets in once the stellar X-ray and extreme-UV (EUV) incident flux is strong enough to greatly heat the planet upper atmosphere [16,17,18,19,20,21]. As a result, the atmosphere starts to dynamically expand pushing the upper atmosphere beyond the planet's Roche lobe or protecting magnetosphere [22], leading to a consistent mass-loss [5]. Heavy atoms (e.g., C, O, Mg, and Fe; [23,15,24,25,26,27]) have also been detected in the exosphere of these planets, proving that heavy elements present in the lower atmosphere can be dragged up in the evaporating exosphere.

Despite these general similarities, the HST data led to the detection/prediction of several different unexpected phenomena: the presence of a cometary tail of planet evaporated material in the case of HD\,209458\,b and HD\,189733\,b [28], and, for WASP-12\,b, the detection of an early ingress, likely caused by a dense bow-shock ahead of the planet [15,29,30] or of a plasma torus originating from an active satellite [31], and of a circumstellar cloud of planet evaporated material [25,32]. In addition, scarce observational hints collected so far with HST provide tantalizing indications that surprisingly large signatures of atmospheric escape could be retrieved even from moderately irradiated gas giants [12] and Neptune-mass planets [13].

Atmospheric evaporation is the only major common denominator gathered so far from the observations, and models predict that many close-in planets should be affected by evaporation [33,34,35,36]. The reality is that it is completely obscure to us how the wealth of phenomena discovered so far depend upon the system parameters: stellar mass\,/\,temperature\,/\,metallicity\,/\,activity, planet mass\,/\,radius\,/\,lower atmospheric structure\,/magnetic field, orbit eccentricity, and semi-major axis. A large number of models has been made to make such predictions, but we completely lack the observational material necessary to constrain and improve them.

The small number of observed systems does not allow us to understand the phenomenology of the planet upper atmosphere and plasma environment. This is fundamental to further improve our general understanding of planet atmosphere, formation, and evolution.

\bigskip

{\bf \large \noindent Science goals of a UV observational parameter study}

\smallskip

\noindent The solution is to collect a treasury data archive of UV transit observations for a sample of systems selected on the basis of their stellar, planet, and orbital characteristics in order to perform an observational parameter study. There are now a sufficient number and variety of planetary systems allowing us to perform such a study, and HST is the only telescope capable of collecting the necessary observations for many years to come. The possible legacy impact of such a data archive is shown for example by the continuous and extensive use of the STIS observations of HD\,209458\,b obtained in 2002 [2] that over the years have been analysed more than a dozen times to extract information about the planet, its plasma environment, and host-star properties.

\smallskip
{\bf \noindent Planet evolution via evaporation} $-$ Evaporation has a major impact on the evolution of a planet atmosphere. Since a phenomenon can usually be studied best where it is strongest, UV transit light curves of a variety of close-in planets with different characteristics would give us the best possible opportunity to study planet evaporation. This is key to understand for example the gap observed by Kepler for intermediate-mass planets with a short orbital separation [37,11,38]. It is crucial to now use HST to understand the diversity of nearby exoplanets that will be discovered with current and future space missions, such as K2, CHEOPS, TESS, and PLATO.  

\smallskip
{\bf \noindent Circumplanetary environment} $-$ Only with further UV transit observations of different close-in planets we can constrain the various models (e.g., [39,29,30,31]) of the early-ingress phenomenon, discovered for WASP-12\,b.

The WASP-12\,b observations led also to the detection of a thick circumstellar cloud formed of material lost by the planet. There are both observational [40] and theoretical [32] works suggesting that this is a common feature of many close-in planets. Because of the different effects these clouds have on different stellar emission lines, far-UV observations would allow us to directly detect the clouds and to characterise their physical properties. 

\smallskip
{\bf \noindent Host star characterisation and effects on exoplanets} $-$ The UV properties of late-type stars are still poorly constrained and UV planet transit observations would provide important information on stellar activity (i.e., UV intensity and variability) and its evolution, which has an important impact on planet evolution as well [41].

Stellar winds play a fundamental role in shaping the formation and evolution of planetary systems, starting with the protoplanetary disk phase [42], but, because of their optically thin nature, they are exceptionally hard to observe and study (e.g., [43]). Charge exchange between stellar wind protons and planetary exosphere neutral hydrogen atoms leads to planet evaporation [44]; constraining stellar winds is therefore crucial to understand planet evaporation as well. In addition, the far-UV radiation has an impact on the atmospheric photochemistry of the planet [45] and allows one to empirically estimate the stellar EUV flux responsible for the heating of the planet atmosphere [46].

UV transit observations of close-in planets give us the unique opportunity to constrain the wind physical properties (temperature, density, and velocity), particularly around the crucial transition region where the wind is accelerated, which is otherwise possible only for the Sun (e.g., [28,47]).

\smallskip
{\bf \noindent Planet atmospheric structure and composition} $-$ The analysis of transit light curves obtained at wavelengths of specific UV spectral lines allows one to infer the temperature, pressure, and dynamical structure of an exoplanet upper atmosphere. As different wavelengths probe different planet atmospheric depths (e.g., [48]), observations obtained in the UV (HST), optical (HST), and infrared (HST/Spitzer/JWST) give us the unique opportunity to consistently constrain models of a whole exoplanet atmosphere, currently possible just for a few Solar System planets and HD\,209458\,b. A set of UV transit observations together with future JWST data will provide the necessary set of observational constraints for the advance of planet atmospheric models. 

A large theoretical and observational effort is being put into trying to understand the chemical composition of extrasolar planets and its importance in governing structures (e.g., planetary mass and radius). Various atoms and molecules have been so far discovered through transmission/emission spectroscopy at different wavelengths. UV transit observations allow one to constrain the abundance of atoms in the upper atmosphere, an information that would be extremely important to best model the chemistry of the lower atmosphere in order to interpret JWST data in the most accurate way.

Optical transit observations revealed that high altitude clouds are a common feature in the atmosphere of close-in planets (e.g., [49]). Their presence is recognised by a planet radius vs. wavelength relation that follows a Rayleigh/Mie scattering law. The STIS observations obtained to date concentrate in the optical and for many planets only the addition of data-points in the near-UV would allow one to constrain the chemical composition, particle size, and altitude of the clouds [50].

\smallskip
{\bf \noindent Planet magnetic moment} $-$ The magnetic field of an exoplanet plays a significant role in the evolution of a planetary atmosphere, but the magnetic properties of exoplanets are still unknown. The detection of an exoplanet magnetic field has been attempted several times, mostly with radio observations [51], but with little success. UV observations, combined with modelling, can instead provide estimates of the planet magnetic moment [29,31,47], which plays an important role in the escape processes, and hence on the whole atmospheric structure [52,53]. 


\smallskip
{\bf \noindent Planet habitability} $-$ In the early stages of the evolution of planetary systems, planets in the habitable zone will experience thermospheric expansion. As a result, evaporation is a key process in the determination of the actual habitability of planets orbiting in the habitable zone [54,55]. It is only with further UV observations that we can constrain the numerous planet atmosphere evolution models that are currently used to study and determine habitability.

\bigskip

{\bf \large \noindent Summary}

\smallskip

\noindent Exoplanet science is now in its full expansion, particularly after the CoRoT and Kepler space missions that led us to the discovery of thousands of extra-solar planets. The last decade has taught us that UV observations play a major role in advancing our understanding of planets and of their host stars, but the necessary UV observations can be carried out \emph{only} by HST, and this is going to be the case for many years to come. It is therefore crucial to build a treasury data archive of UV exoplanet observations formed by a dozen ``golden systems'' for which observations will be available from the UV to the infrared. Only in this way we will be able to fully exploit JWST observations for exoplanet science, one of the key JWST science case.

\bigskip

\begin{scriptsize}

{\bf \noindent References}: 
[1]\,Charbonneau et al.\,2002, ApJ, 568, 377;
[2]\,Vidal-Madjar et al.\,2003, Nature, 422, 143;
[3]\,Koskinen et al.\,2010, ApJ, 723, 116;
[4]\,Lopez \& Fortney 2013, ApJ, 776, 2;
[5]\,Lecavelier des Etangs et al.\,2004, A\&A, 418, L1;
[6]\,Leitzinger et al. 2011, P\&SS, 59, 1472;
[7]\,Kurokawa \& Kaltenegger 2013, MNRAS, 433, 3239;
[8]\,Li et al.\,2010, Nature, 463, 1054;
[9]\,Davis \& Wheatley 2009, MNRAS, 396, 1012;
[10]\,Howard et al. 2012, ApJS, 201, 15;
[11]\,Beaug{\'e} \& Nesvorn{\'y} 2013, ApJ, 763, 12;
[12]\,Ehrenreich et al. 2012, A\&A, 547, A18;
[13]\,Kulow et al. 2014, ApJ, 786, 132;
[14]\,Lecavelier Des Etangs et al.\,2010, A\&A, 514, A72;
[15]\,Fossati et al.\,2010, ApJL, 714, L222;
[16]\,Yelle, 2004, Icarus, 170, 167;
[17]\,Garc{\'{\i}}a Mu{\~n}oz 2007, P\&SS, 55, 1426;
[18]\,Koskinen et al. 2007, Nature, 450, 845;
[19]\,Murray-Clay et al. 2009, ApJ, 693, 23;
[20]\,Koskinen et al. 2013, Icarus, 226, 1678;
[21]\,Koskinen et al. 2013, Icarus, 226, 1695;
[22]\,Lichtenegger et al. 2010, Icarus, 210, 1;
[23]\,Vidal-Madjar et al.\,2004, ApJ, 604, L69;
[24]\,Linsky et al.\,2010, ApJ, 717, 1291;
[25]\,Haswell et al.\,2012, ApJ, 760, 79;
[26]\,Ben-Jaffel \& Ballester\,2013, A\&A, 553, A52;
[27]\,Vidal-Madjar et al.\,2013, A\&A, 560, A54;
[28]\,Bourrier \& Lecavelier des Etangs 2013, A\&A, 557, A124;
[29]\,Vidotto et al.\,2010, ApJL, 722, L168;
[30]\,Bisikalo et al.\,2013, ApJ, 764, 19;
[31]\,Ben-Jaffel \& Ballester, 2014, ApJL, 785, L30;
[32]\,Fossati et al.\,2013, ApJL, 766, L20;
[33]\,Lecavelier des Etangs 2007, A\&A, 461, 1185;
[34]\,Ehrenreich \& D{\'e}sert 2011, A\&A, 529, A136;
[35]\,Lopez et al. 2012, ApJ, 761, 59;
[36]\,Owen \& Wu 2013, ApJ, 775, 105;
[37]\,Mordasini et al.\,2012, A\&A, 547, A112;
[38]\,Jin et al.\,2014, ApJ, 795, 65;
[39]\,Lai et al.\,2010, ApJ, 721, 923;
[40]\,Lanza\,2014, A\&A, 572, L6;
[41]\,Linsky \& G\"udel 2015, ASSL, 411, 3;
[42]\,France, et al.\,2011, ApJ, 729, 7;
[43]\,Wood et al. 2005, ApJL, 628, L143;
[44]\,Holmstr{\"o}m et al. 2008, Nature, 451, 970;
[45]\,Hu et al. 2012, ApJ, 761, 166;
[46]\,Linsky et al. 2014, ApJ, 780, 61;
[47]\,Kislyakova et al.\,2014, Science, 346, 981;
[48]\,Fossati et al. 2015, ASSL, 411, 59;
[49]\,Sing et al.\,2015, MNRAS, 446, 2428;
[50]\,Wakeford \& Sing\,2014, A\&A, 573, A122;
[51]\,Sirothia et al. 2014, A\&A, 562, A108;
[52]\,Trammell et al. 2011, ApJ, 728, 152;
[53]\,Owen \& Adams 2014, MNRAS, 444, 3761;
[54]\,Buccino et al. 2006, Icarus, 183, 491;
[55]\,France et al. 2013, ApJ, 763, 149.
\end{scriptsize}

\end{document}